\newcommand{\ee}{\mathrm{e}}
\newcommand{\xperp}{\mathbf x}
\newcommand{\yperp}{\mathbf y}

\newcommand{\bperp}{\mathbf b}
\newcommand{\elrf}{\epsilon_\mathrm{LRF}}

\documentclass[a4paper,11pt]{article}
\usepackage{pos}
\usepackage{amsmath}
\usepackage{physics}
\usepackage{booktabs}

\title{Charged particle multiplicity in pp-collisions from the dilute Glasma}
\author[a]{Andreas Ipp}
\author*[a]{Markus Leuthner}
\author[a]{David I. M\"uller}
\author[b]{S\"oren Schlichting}
\author[a]{Kayran Schmidt}
\author[c,d]{Pragya Singh}

\affiliation[a]{Institute for Theoretical Physics, TU Wien,\\
Wiedner Hauptstraße  8-10/136, A-1040 Vienna, Austria }

\affiliation[b]{Fakult\"at f\"ur Physik, Universit\"at Bielefeld,\\
D-33615 Bielefeld, Germany}

\affiliation[c]{Department of Physics, University of Jyv\"askyl\"a,\\
FI-40014 Jyv\"askyl\"a, Finland}

\affiliation[d]{Helsinki Institute of Physics, University of Helsinki,\\
FI-00014 Helsinki, Finland}

\emailAdd{ipp@hep.itp.tuwien.ac.at}
\emailAdd{mleuthner@hep.itp.tuwien.ac.at}
\emailAdd{dmueller@hep.itp.tuwien.ac.at}
\emailAdd{sschlichting@physik.uni-bielefeld.de}
\emailAdd{kschmidt@hep.itp.tuwien.ac.at}
\emailAdd{prasingh@jyu.fi}

\abstract{Proton-proton collisions are studied in the dilute Glasma framework. Compared to experimental multiplicity distributions, the dilute Glasma underestimates large multiplicity events. We show how event-by-event fluctuations of the saturation momentum $Q_s$ can help repair the multiplicity distribution. Furthermore, we discuss a hot spot model for protons and also find improvements in the multiplicity histograms.}

\FullConference{The XVIth Quark Confinement and the Hadron Spectrum Conference (QCHSC24)\\
 19-24 August, 2024\\
 Cairns Convention Centre, Cairns, Queensland, Australia\\}

\begin{document}

\renewcommand{\hookAfterAbstract}{%
\par\bigskip\bigskip
\textsc{ArXiv ePrint}:
\href{https://arxiv.org/abs/XXXXXX}{XXXXXX}}

\maketitle

\section{Introduction}
The earliest stage in a collision of atomic nuclei at high energies is called the Glasma \cite{
%Kovner:1995ja, Kovner:1995ts, Krasnitz:1999wc, Krasnitz:2000gz,
Lappi:2006fp}. It is a classical state characterized by strong coherent color fields and the precursor to the quark-gluon plasma \cite{
%Shuryak:2004cy, Busza:2018rrf,
Gelis:2021zmx}. The Glasma arises in an effective description of high energy QCD called the Color Glass Condensate (CGC) \cite{
%Iancu:2003xm, McLerran:2008es,
Gelis:2010nm%
%, Gelis:2012ri
}. The hard partons within the colliding nuclei are modeled as classical color charges, which act as sources for the soft partons described as a classical gauge field in the classical Yang-Mills equations. The additional gauge field produced in the collision of two such nuclei is the Glasma. Most analyses of the Glasma make use of the boost invariant approximation, where nuclei are considered to move at the speed of light and are not allowed to have nontrivial longitudinal structure. Consequently, all information about the rapidity dependence of observables is lost. (3+1)D simulations of the Glasma \cite{
%Gelfand:2016yho, Ipp:2017lho, Ipp:2018hai,
Ipp:2020igo, Schlichting:2020wrv, Matsuda:2023gle%
%, Matsuda:2024moa, Matsuda:2024mmr
} incorporate 3D initial conditions and can resolve the longitudinal structure of the Glasma, but are computationally expensive. The (3+1)D dilute Glasma \cite{Ipp:2021lwz, Ipp:2022lid, Ipp:2024ykh, Leuthner:2025vsd} was recently introduced as a new semianalytic framework that allows the computation of observables in the (3+1)D Glasma at arbitrary times assuming weak sources. In these proceedings, we apply the dilute Glasma framework to the study of proton-proton (pp) collisions.
\section{The dilute Glasma}
A single nucleus or nucleon is described in the CGC framework by the $\mathfrak{su}(3)$-valued color current
\begin{align}
    \mathcal J^\mu_{A/B}(x^\pm, \xperp) = \delta^\mu_{\mp}\rho_{A/B}(x^\pm, \xperp),
\end{align}
where the labels $A$ and $B$ correspond to the two collision partners \cite{Ipp:2024ykh}. Due to their relativistic motion, the color currents only exhibit a single vector component and depend on a single lightcone coordinate $x^+$ or $x^-$. Both are allowed to depend on the transverse coordinates $x$ and $y$, collectively denoted as $\xperp$.
Assuming covariant gauge, $\partial_\mu \mathcal A^\mu_{A/B}(x^\pm, \xperp)=0$, the single-particle gauge fields are obtained from the respective charges by solving the transverse Poisson equation
\begin{align}
    \label{eq:tranverse_poisson}
    -\nabla_\perp^2\mathcal A^\mp_{A/B}(x^\pm, \xperp) = \rho_{A/B}(x^\pm, \xperp).
\end{align}
The total gauge field in the collision system can be written as
\begin{align}
    A^\mu(x^+,x^-,\xperp) = \mathcal A_A(x^+,\xperp) + \mathcal A_B(x^-, \xperp) + a^\mu(x^+,x^-,\xperp).
\end{align}
The term $a^\mu(x^+,x^-,\xperp)$ captures the nonlinearities due to the interaction of the collision partners. It is localized in the causal future of the interaction region and it is the Glasma field that we are interested in.
Finding an expression for $a^\mu(x^+,x^-,\xperp)$ requires solving the Yang-Mills equations for the full collision problem, which has not been achieved analytically. The dilute Glasma framework yields closed-form expressions for $a^\mu(x^+,x^-,\xperp)$ under the assumption of weak sources, i.e., ignoring subleading contributions in a power series expansion in $\mathcal J_{A/B}$. At this order, the field strength tensor becomes Abelian, $f^{\mu\nu} = \partial^\mu a^\nu - \partial^\nu a^\mu$.
Compact closed-form expressions for the field strength tensor of the dilute Glasma were presented in \cite{Ipp:2024ykh}.
%\begin{align}
%    f^{+-}(x) &= -\frac{g}{2\pi}\int_{-\infty}^\infty \dd{\eta'}\intop_\vperp V(x,\eta',\vperp),\label{eq:fpm}\\
%    f^{\pm i}(x) &= \frac{g}{2\pi} \int_{-\infty}^\infty\dd{\eta'} \intop_\vperp \left(V^{ij}(x,\eta',\vperp)\mp\delta^{ij}V(x,\eta',\vperp)\right)w^j\frac{\ee^{\pm \eta'}}{\sqrt{2}},\\
%    f^{ij}(x) &= -\frac{g}{2\pi}\int_{-\infty}^\infty \dd{\eta'}\intop_\vperp V^{ij}(x,\eta',\vperp).\label{eq:fij}
%\end{align}
From the field strength tensor, gauge invariant observables such as the energy-momentum tensor can be computed.
\section{Proton-proton collisions in the dilute Glasma}
The dilute Glasma framework can be applied to collisions of relativistic protons as explored in \cite{Leuthner:2025vsd}. We sample color charges from a Gaussian probability functional with vanishing 1-point function,
\begin{align}
    \langle \rho_{A/B}^a(x^\pm, \xperp)\rangle = 0.
\end{align}
The 2-point function is \cite{Leuthner:2025vsd}
\begin{align}
\label{eq:correlator_single_nucleon}
    &\langle \rho^a_{A/B}(x^\pm,\xperp)\rho^b_{A/B}(y^\pm, \yperp)\rangle= g^2\mu^2\delta^{ab}\frac{1}{\sqrt{2\pi}s_l}\ee^{-\frac{(x^\pm+y^\pm)^2}{8s_l^2}}\ee^{-
    \vphantom{\frac{(x^\pm+y^\pm)^2}{8s_l^2}}
    \frac{(\xperp+\yperp)^2}{8s^2}}\frac{1}{\sqrt{2\pi}\xi}\ee^{-
    \vphantom{\frac{(x^\pm+y^\pm)^2}{8s_l^2}}
    \frac{(x^\pm-y^\pm)^2}{2\xi^2}}\delta^{(2)}(\xperp-\yperp),
\end{align}
where the parameter $s$ governs the (transverse) size of the proton, which has a Gaussian shape. Also, $s_l = s/(\sqrt{2}\gamma)$ is the corresponding longitudinal size. The parameter $\xi$ is a correlation length and sets the size of fluctuations in the longitudinal direction. Latin indices denote coefficients with respect to a basis of $\mathfrak{su}(3)$.
\begin{figure}[t]
    \centering
\includegraphics[width=0.55\linewidth]{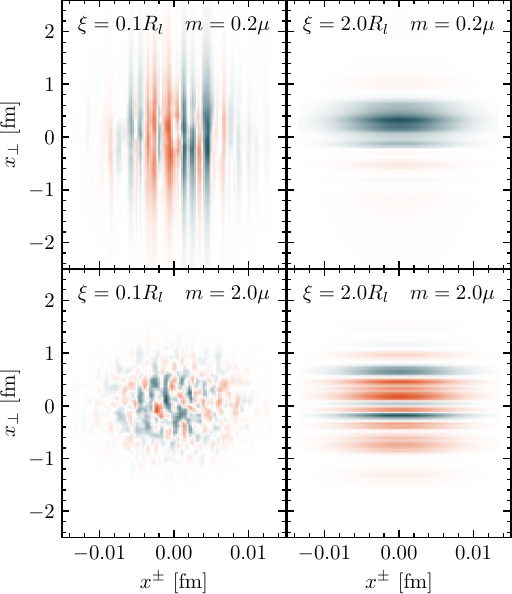}
    \caption{Gauge field component $\mathcal A^{\mp, a}$ of a single proton for different values of the infrared regulator $m$ and correlation length $\xi$. The transverse coordinate $x_\perp$ stands for either transverse direction. Orange regions correspond to positive values and blue regions correspond to negative values.}
    \label{fig:proton}
\end{figure}
In Fig.~\ref{fig:proton} we show one color component of the gauge field of a single proton under this model. Note that the infrared (IR) regulator $m$, which is employed when solving the Yang-Mills equations for a single proton, governs the size of fluctuations in the transverse direction. Conversely, the correlation length parameter $\xi$ governs the scale of fluctuations in longitudinal direction. In the following, we keep $\xi$ at the maximum allowed value $\xi =2s_l$.\par
To reproduce the experimentally observed proton-proton cross-section we introduce an impact parameter-dependent interaction probability \cite{dEnterria:2010xip}.
The enveloping transverse profile of a single proton modeled after \eqref{eq:correlator_single_nucleon} and normalized in the transverse plane is
\begin{align}
\label{eq:nucleon_thickness}
    t(\xperp) = \frac{1}{2\pi s^2}\exp\left(-\frac{\xperp^2}{2s^2}\right).
\end{align}
For two protons separated in the transverse direction by the impact parameter $\bperp$ we define the overlap function
\begin{align}
\label{eq:overlap_function}
    \Theta(b) = \int \dd[2]{\xperp}\,t(\xperp+\bperp/2)t(\xperp-\bperp/2) = \frac{1}{4\pi s^2}\exp\left(-\frac{b^2}{4 s^2}\right),
\end{align}
where $b$ = $|\bperp|$.
The mean number of parton-parton collisions between two protons can then be expressed as $N_{gg} = N_g^2 \sigma_{gg}\Theta(b)$. Here, 
$N_g$ is the parton number per nucleon and $\sigma_{gg}$ is the parton-parton cross-section.
Assuming that parton-parton collisions are independent processes, their statistics should follow a Poisson distribution. The differential probability of one or more parton-parton interactions is then
\begin{align}
\label{eq:differential_probability}
\dv[2]{P}{\bperp}(b) = \frac{1-\exp\left(-N_g^2\sigma_{gg} \Theta(b)\right)}{\int \dd[2]{\bperp}\left(1-\exp\left(-N_g^2\sigma_{gg} \Theta(b)\right)\right)}.
\end{align}
The denominator of this expression is the effective proton-proton cross-section $\sigma_{NN}$.
We match $\sigma_{NN}$ to the experimental inelastic proton-proton cross-section, for which the interpolation formula 
\cite{McLerran:2015qxa}
\begin{align}
\label{eq:sigma_inel_interpolation}
    \sigma_{NN} &= \left[2.52 + 0.005\,\ln\left(\frac{\sqrt{s}}{\mathrm{GeV}}\right)+0.056\,\ln(\frac{\sqrt{s}}{\mathrm{GeV}})^2+4.52\,\left(\frac{\sqrt{s}}{\mathrm{GeV}}\right)^{-0.9}+3.38\,\left(\frac{\sqrt{s}}{\mathrm{GeV}}\right)^{-1.1}\right]\,\mathrm{fm}^2
\end{align}
has been found.
This fixes the parameter combination $N_g^2 \sigma_{gg}$. No other knowledge about $N_g$ and $\sigma_{gg}$ is required.
\section{Numerical results}
\begin{figure}[t]
    \centering
\includegraphics[width=\linewidth]{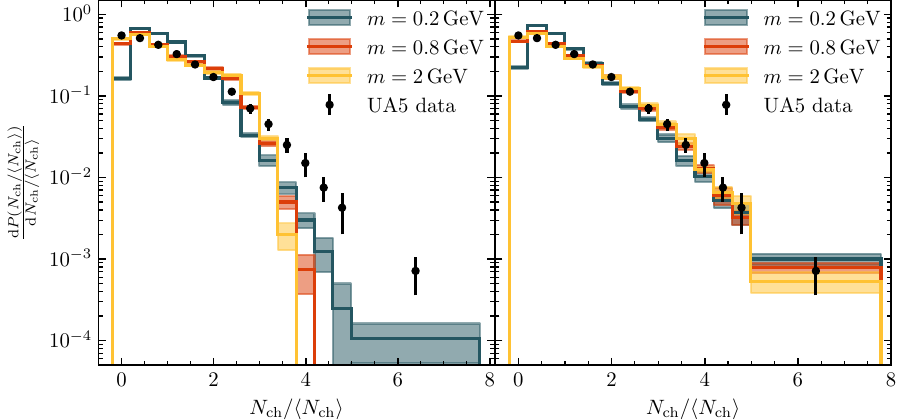}
    \caption{Histogram of charged particle multiplicity $N_\mathrm{ch}$ for proton-proton collisions in the dilute Glasma at $\sqrt{s}=200\,\mathrm{GeV}$. $10^4$ collisions were simulated at each of three different values for the IR cutoff $m$ and contrasted with experimental $N_\mathrm{ch}$ distributions in the $-0.5 < \eta_s < 0.5$ rapidity range of proton-antiproton collisions by the UA5 collaboration \cite{UA5:1988gup, hepdata.15457.v1/t3}. In the left diagram, no additional modifications to the dilute Glasma framework were used. For the right plot, additional event-by-event fluctuations of $Q_s$ were employed. Error bands on simulation results were obtained via the jackknife method.}
    \label{fig:hist_unscaled_scaled}
\end{figure}

We fix the parameter $s$ in the thickness function \eqref{eq:nucleon_thickness} to $s=0.395\,\mathrm{fm}$, which was fitted to HERA data in \cite{Rezaeian:2012ji}. To sample values of the impact parameter $b$ from the distribution \eqref{eq:differential_probability} we use the slightly different value $s=0.358\,\mathrm{fm}$ taken from \cite{Mantysaari:2016jaz} instead. With these parameters, we sample $10^4$ proton-proton collision events at $\sqrt{s}=200\,\mathrm{GeV}$ and different values of the IR cutoff $m$. The resulting local rest frame energy density $\elrf$ of the Glasma at mid-rapidity ($\eta_s = 0$) is taken as a proxy for the charged particle multiplicity produced in the collision and detected in the most central bin around $\eta_s = 0$. We rescale these results by the average multiplicity and present them in the left diagram of Fig.~\ref{fig:hist_unscaled_scaled}.
Regardless of the value of the IR cutoff, events with large multiplicity are underrepresented in the dilute Glasma, consistent with results from IP-Glasma simulations \cite{Schenke:2013dpa, McLerran:2015qxa}. In the same context, modifications were suggested to improve the multiplicity histogram and recover the high multiplicity events.
\subsection{\texorpdfstring{Fluctuations of $Q_s$}{Fluctuations of Qs}}
Event-by-event fluctuations of the ratio between the saturation momentum $Q_s$ and the model parameter $g^2\mu$ as described in \cite{Schenke:2013dpa} were later reformulated into fluctuations of $Q_s$ for each colliding proton drawn from the distribution \cite{McLerran:2015qxa}
\begin{align}
\label{eq:Q_s_dist}
    P\left(\ln(Q_s^2/\langle Q_s^2\rangle )\right)=\frac{1}{\sqrt{2\pi}\sigma}\exp\left( -\frac{\ln^2(Q_s^2/\langle Q_s^2\rangle)}{2\sigma^2}\right).
\end{align}
In the context of the dilute Glasma, this can be straightforwardly translated to fluctuations of the color charges of individual protons with a factor $\sqrt{Q_s^2/\langle Q_s^2\rangle}$ drawn from the distribution \eqref{eq:Q_s_dist}. The parameter $\sigma$ varies weakly with collision energy. A value of $\sigma \approx 0.5$ has been used in the literature \cite{McLerran:2015qxa%,
%Bzdak:2015eii
}.
We do not allow for an energy dependence of $\sigma$ and fix its value at a given IR cutoff $m$ by minimizing the Kullback-Leibler divergence
\begin{align}
    D_\mathrm{KL}=\sum_i P_i \ln \left(\frac{P_i}{Q_i}\right)
\end{align}
between the experimental multiplicity distribution ($P$)  and the one obtained from the dilute Glasma ($Q$)\footnote{As argued in \cite{Leuthner:2025vsd}, we do not consider the lowest multiplicity bin when computing $D_\mathrm{KL}$.}.
%\begin{figure}[t]
%    \centering
%\includegraphics[width=0.55\linewidth]{Proceedings_Markus/KL.pdf}
%    \caption{Kullback-Leibler divergence between multiplicity distributions of proton-proton collisions at $\sqrt{s}=200\,\mathrm{GeV}$ in the dilute Glasma and experimental results. The simulations were run with different values of $\sigma$ to find the optimal value for each choice of IR cutoff $m$. Experimental data are obtained from the UA5 collaboration \cite{UA5:1988gup, hepdata.15457.v1/t3}. The first multiplicity bin is not included in the evaluation.}
%    \label{fig:KL}
%\end{figure}
$D_\mathrm{KL}$ is minimal for the optimal values $\sigma_\mathrm{opt}$ shown in Tab.~\ref{tab:sigma_opt}.
\begin{table}[htbp]
  \centering
    \begin{tabular}{r|lll}
    $m$     & $0.2\,\mathrm{GeV}$ & $0.8\,\mathrm{GeV}$ & $2\,\mathrm{GeV}$ \\
    \midrule
    $\sigma_\mathrm{opt}$ & 0.26     & 0.22     & 0.21 \\
    \end{tabular}%
  \caption{Optimal values $\sigma_\mathrm{opt}$ for the width of the distribution \eqref{eq:Q_s_dist} at $\sqrt{s}=200\,\mathrm{GeV}$.}%
  \label{tab:sigma_opt}%
\end{table}
%\begin{figure}[t]
%    \centering
%\includegraphics[width=0.55\linewidth]{Proceedings_Markus/200_hist_Qs_scaled.pdf}
%    \caption{Multiplicity histogram for proton-proton collisions in the dilute Glasma with additional event-by-event fluctuations of $Q_s$. $10^4$ collisions at $\sqrt{s}=200\,\mathrm{GeV}$ were simulated at mid-rapidity and compared to data for proton-antiproton collisions in the $-0.5 < \eta_s < 0.5$ rapidity range by the UA5 collaboration \cite{UA5:1988gup, hepdata.15457.v1/t3}. Error bands on simulation results were obtained via the jackknife method.}
%    \label{fig:hist_Qs_scaled}
%\end{figure}
Note that these values of $\sigma_\mathrm{opt}$ are about half as large as those previously used in the literature.
In the right diagram of Fig.~\ref{fig:hist_unscaled_scaled} we show the resulting multiplicity histogram for proton-proton collisions in the dilute Glasma at $\sqrt{s}=200\,\mathrm{GeV}$ employing fluctuations of $Q_s$ drawn from the distribution \eqref{eq:Q_s_dist} with $\sigma_\mathrm{opt}$ as shown in Tab.~\ref{tab:sigma_opt}. The multiplicity distribution has improved significantly and now matches the experimental distribution, especially for the two larger values of $m$.
\subsection{Hot spot model}
\begin{figure}[t]
    \centering
\includegraphics[width=\linewidth]{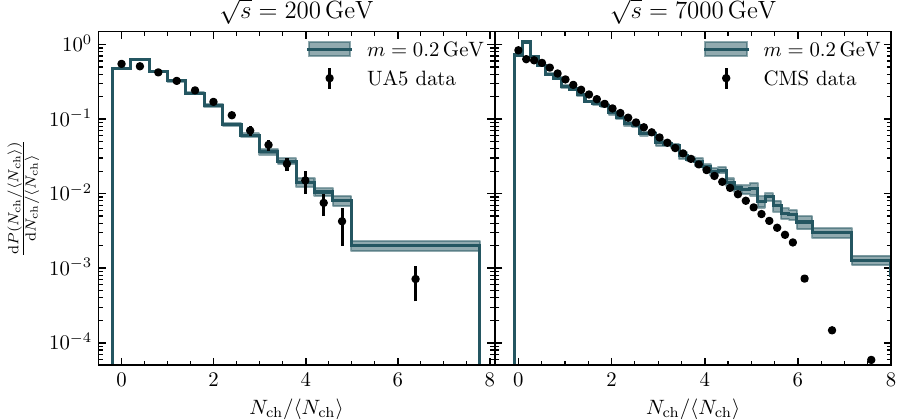}
    \caption{Multiplicity histograms for proton-proton collisions in the dilute Glasma employing the hot spot model. Simulation results are contrasted with experimental results by the UA5 collaboration \cite{UA5:1988gup, hepdata.15457.v1/t3} at $200\,\mathrm{GeV}$ (left) and the CMS collaboration \cite{CMS:2010qvf, hepdata.57909.v1/t12} at $7000\,\mathrm{GeV}$ (right). Error bands on simulation results were obtained via the jackknife method.}
    \label{fig:n_v_hist}
\end{figure}

Another approach to improve the multiplicity histogram is to consider protons that consist of 3 individual hot spots, representing the valence quarks. We model hot spots as scaled-down versions of the protons described by \eqref{eq:correlator_single_nucleon}. Following \cite{Mantysaari:2022ffw}, the size of individual hot spots is $0.116\,\mathrm{fm}$ and they are scattered with a Gaussian distribution of width $s=0.416\,\mathrm{fm}$ around the virtual proton center. The actual proton center, used for impact parameter calculations, is then taken as the average of the 3 hot spot coordinates. We simulate $10^4$ proton-proton collisions at $\sqrt{s}=200\,\mathrm{GeV}$ and $2.5\times 10^4$ collisions at $\sqrt{s}=7000\,\mathrm{GeV}$ in the dilute Glasma employing the hot spot model. The resulting multiplicity histograms are shown in Fig.~\ref{fig:n_v_hist}.
There is a clear improvement compared to the distribution in the left diagram of Fig.~\ref{fig:hist_unscaled_scaled} for the histogram at $\sqrt{s}=200\,\mathrm{Gev}$. Results at $\sqrt{s}=7000\,\mathrm{GeV}$ are compared to CMS data \cite{CMS:2010qvf, hepdata.57909.v1/t12}. They agree for most multiplicity bins, and only for the highest multiplicity bins, the dilute Glasma now overestimates the frequency of these events.
%%%
\section{Conclusion}
We showed that proton-proton collisions in the dilute Glasma framework require additional modifications to reproduce experimentally measured multiplicity distributions. In particular, high multiplicity events are underrepresented in the dilute Glasma. Fluctuations of the saturation momentum $Q_s$ are interpreted as event-by-event rescalings of the proton color charges in the dilute Glasma. These fluctuations, if drawn from a distribution with the right parameter $\sigma$, can massively improve the multiplicity histograms. As an alternative approach, we also improved the proton model by allowing for 3 individual hot spots in the proton. This additional ``lumpiness'' also improves the multiplicity distribution and even slightly overestimates the large multiplicity events.\par
These results serve as a stepping stone towards calculations of observables in collisions of protons and nuclei in the dilute Glasma. Event plane decorrelations were studied experimentally in Pb-p collisions \cite{CMS:2015xmx} and could be reproduced in dilute Glasma simulations. Such small collision systems (as compared to nucleus-nucleus collisions previously studied in the dilute Glasma) can be efficiently treated numerically and, as shown in these proceedings, the additional substructure in the initial conditions introduced by the hot spot model is likely to play an important role.
\acknowledgments
\hyphenation{Deut-sche}
\hyphenation{For-schungs-ge-mein-schaft}
\hyphenation{Dok-to-rats-kol-leg}
ML, DM, and KS are supported by the Austrian Science Fund FWF No.\ P34764.
SS is supported by Deutsche Forschungsgemeinschaft (DFG, German Research Foundation), CRC-TR 211 \lq Strong-interaction matter under extreme conditions\rq – project number 315477589 – TRR 211.
PS is supported by the Academy of Finland, the Centre of Excellence in Quark Matter (project 346324), and the European Research Council, ERC-2018-ADG-835105 YoctoLHC.
The computational results have been achieved in part using the Vienna Scientific Cluster (VSC).
\bibliographystyle{JHEP}
\bibliography{references}

\end{document}